# High mobility two-dimensional hole system in GaAs/AlGaAs quantum wells grown on (100) GaAs substrates

M. J. Manfra, L. N. Pfeiffer, K. W. West, R. de Picciotto, and K. W. Baldwin

Bell Laboratories, Lucent Technologies

600 Mountain Ave., Murray Hill, NJ 07974

We report on the transport properties of a high mobility two-dimensional hole system (2DHS) confined in GaAs/AlGaAs quantum wells grown molecular-beam epitaxy on the (100) surface of GaAs. The quantum wells are modulation-doped with carbon utilizing a novel resistive filament source. At T=0.3K and carrier density $p=1 \times 10^{11} cm^{-2}$, a mobility of $10^6 cm^2/Vs$ is achieved. At fixed carrier density $p=10^{11} cm^{-2}$, the mobility is found to be a non-monotonic function of the quantum well width. The mobility peaks at $10^6 cm^2/Vs$ for a 15nm well and is reduced for both smaller and larger well widths for these (100) samples. The mobility anisotropy is found to be small. Mobility along (0$\underline{1}$1) is approximately 20% higher than along the (011) direction. In addition, the low temperature carrier density is found to have low sensitivity to light. The hole density increases by only ~10% after exposure to red light at T=4.2K. In structures designed for a lower carrier density of $3.6 \times 10^{10} cm^{-2}$, a mobility of $800,000 cm^2/Vs$ is achieved at T=15mK.

The first studies of two-dimensional hole systems (2DHSs) in GaAs/AlGaAs heterostructures utilized the (100) orientation of GaAs with beryllium (Be) incorporated as the acceptor [1-4]. Due to the increased sample disorder resulting from the high diffusion constant of Be in GaAs, hole mobilities for Be modulation-doped heterostructures are typically limited to ~$10^5$cm$^2$/Vs. As a result, much work in recent years has focused on the study of high mobility 2DHS grown on the 311A orientation of GaAs [5-7]. In the 311A orientation, silicon (Si), which has a diffusion constant 100 times smaller than that of Be [8], can be incorporated on the arsenic site and act as an acceptor. Due to the resulting lower disorder associated with the sharper Si impurity distribution, the 2DHS mobility can be as high as $10^6$cm$^2$/Vs [9-11]. While the high mobility for the 311A hole gas is quite impressive, several limitations have also been recognized. The reduced symmetry of the 311A orientation results in a more complicated band structure than the (100) orientation. In addition, a surface corrugation unique to 311A produces a large mobility anisotropy [9]. The mobility along the ($\bar{2}$33) direction can exceed the mobility along the (01$\bar{1}$) direction by a factor of 2 to 4, complicating the interpretation of transport data. The 311A 2DHS density is also very sensitive to exposure to red light, which *reduces* the 2D density from the value initially obtained by cooling to low temperature in the dark. This fact complicates optical studies of the 2DHS and results in a reduction of mobility upon exposure to light. Finally, the literature suggests that the mobility on the 311A orientation is limited by interface roughness scattering [11], which makes the achievement of mobilities greater than $10^6$cm$^2$/Vs very challenging.

In this letter we report on the growth and transport properties of high mobility 2DHSs grown on the (100) orientation of GaAs. We focus on structures that are modulation-doped with carbon using a resistive carbon filament developed at Bell Laboratories. Carbon is chosen as the p-dopant because below 1000ºC its vapor pressure is very low and its diffusion constant in GaAs is similar to Si at typical MBE growth temperatures [8]. The design of our carbon filament is such that efficient doping can be achieved at a filament power of 100 Watts, which is a factor of ~5 to10 lower power consumption than commercially available carbon filament sources [12]. Using this source we have produced carbon-doped 2DHSs with mobility measured in the van der Pauw geometry of $1x10^6 cm^2/Vs$ at a carrier density of $1x10^{11} cm^{-2}$ at T=0.3K. To the best of our knowledge, this result represents the highest mobility yet achieved for a modulation-doped 2DHS on (100) GaAs [13].

The growth sequence of our heterostructures is as follows. A 500nm GaAs buffer layer is initially grown on a GaAs (100) substrate, followed by 200 repeats of a 3nm GaAs/10nm $Al_{0.32}Ga_{0.68}As$ superlattice. The lower barrier of the quantum well is formed by 10nm of $Al_{0.32}Ga_{0.68}As$. A GaAs quantum well (width 5 to 1000nm) is then deposited. The GaAs quantum well is followed with an 80nm undoped layer of $Al_{0.32}Ga_{0.68}As$. At a distance of 80nm from the quantum well, the sample is delta-doped with carbon at the $2.5x10^{12} cm^{-2}$ level. We note that all structures discussed in this paper are single-side modulation doped. The structure is completed with 100nm of $Al_{0.32}Ga_{0.68}As$, followed by a 10nm GaAs capping layer. The entire structure is grown at 640ºC, except for the delta-doping layer, which is typically deposited at 500ºC. This structure reproducibly resulted in a 2DHS density $p=10^{11} cm^{-2}$ at T=0.3K for all investigated quantum well widths. For the lower density samples ($p=3.6 \times 10^{10} cm^{-2}$) also discussed in the text, the Al content of the quantum well barriers is reduced to 16.5%. In this study, we have systematically varied the quantum well width at a given 2DHS density in order to investigate the role of well width on low temperature transport. For structures with $p=10^{11} cm^{-2}$ the quantum well geometry is varied from a single interface down to a 5nm quantum well. For samples with $p=3.6 \times 10^{10} cm^{-2}$ we have grown quantum wells with width 100nm, 30nm and 15nm. A total of 10 different growth runs are discussed in this work.

      The low temperature mobility of the 2DHS is determined in the van der Pauw geometry. Eight In/Zn contacts are annealed around the perimeter of a 4mm by 4mm square. All samples are cooled in the dark to T=0.3K and the mobility averaged over the square is measured. We note that the mobility for samples with $p=10^{11} cm^{-2}$ is largely saturated by 0.3K. Table 1 summarizes the measurements of all 10 runs, including the lower density samples. In order to measure any mobility anisotropy that may exist, we also employed photolithography to fabricate two mutually perpendicular Hall bars on a single piece of wafer 9-8-03.1. One Hall bar was oriented along the (011) direction, the other along the (01̲1) direction. We found only a slight anisotropy in mobility; the mobility along the (01̲1) direction was approximately 20% higher than along the (011) direction. Interestingly, this mobility anisotropy is consistent with the anisotropy observed for the equivalent two-dimensional *electron* system grown on the (100) orientation [14]. This small anisotropy is in sharp contrast with the factor of 2-4 mobility anisotropy found for 311A samples. In addition, we have examined the sensitivity of the 2DHS density to exposure to red light. In contrast with the behavior for the 311A orientation, the 2DHS density was found to *increase* by approximately 10% upon

exposure to red light at T=4.2K over the value obtained by cooling to low temperature in the dark.  This small perturbation to the carrier density suggests that controlled optical studies of high mobility 2DHSs will be possible with these new carbon-doped (100) samples to a degree not afforded by the 311A structures.

It is evident that the mobility of the quantum well structures with $p=10^{11} cm^{-2}$ depends strongly on the width of the quantum well and exhibits a non-monotonic dependence.  Figure 1 illustrates this behavior.  Starting with a single interface, the mobility is found to *increase* with *decreasing* well width down to a well width of 15nm where the maximum mobility of $10^6 cm^2/Vs$ is measured. For well widths below 15nm the carrier mobility decreases rapidly with decreasing well width.  This non-monotonic behavior has no analog in the equivalent n-type structure.  In n-type heterostructures, the electron mobility is largely independent of well width, until the quantum well is less than approximately 25nm where interface roughness scattering produces a dramatic reduction in mobility for further reductions in well width.  Thus the decrease in mobility at very narrow quantum well widths in our p-type structures is not surprising, and is consistent with interface roughness scattering being the dominant mobility limiting scattering mechanism for well widths <15nm.  What is unusual is that the mobility is higher for the 15nm quantum than for the 30nm and 50nm wells and the single interface.  One can ask if this change in mobility is due a real change in the effective scattering time, or perhaps, is it connected with a change in some other material parameter.  We speculate that this behavior is correlated with the well width dependence of the band edge effective mass ($m_b$) in p-type heterostructures [15].  The valence bands of GaAs are non-parabolic due to the mixing of the light and heavy hole bands away from the Brillouin zone center ($k$=0).

As a result, the valence band edge mass can depend on several factors, including the quantum well confining potential and the 2D density. Interestingly, such an effect has been observed for 2DHSs grown in the 311A orientation. Pan *et al*. [15] observed that the band edge effective mass ($m_b$) for 311A structures with density $3.7 \times 10^{10} cm^{-2}$ was reduced from $0.44 m_e$ in a 30nm quantum to $0.19 m_e$ for a 10nm quantum well. Due to the strong quantum confinement in narrow wells, the light hole state and the heavy hole state are pushed further apart in energy, reducing the admixture of the heavy hole and light hole bands away from *k*=0. Consequently, the band edge mass remains lighter. For wider quantum wells, the influence of the confinement potential is not as significant and thus the mixing of the light and heavy hole bands remains strong and results in a higher effective mass. The results of Ref. 15 are consistent with the fact that we observe a higher mobility in our narrower (15nm) quantum well. Evidently for well widths less than 15nm, the influence of interface roughness scattering becomes the dominant effect on mobility and obscures any other well width dependence. In order to clarify the role of effective mass in our (100) structures, experiments are currently underway to use cyclotron resonance to measure the effective mass over a wide range of quantum well widths and density.

    In addition to T=0.3K mobility, we also utilize low temperature magnetotransport measurements in the integer and fractional quantum Hall regime to access the quality of these carbon-doped 2DHSs. Figure 2 displays the transport coefficients $R_{xx}$ and $R_{xy}$ measured at T=20mK for sample 9-8-03.1. The excitation current is limited to ≤1nA. Clearly this sample is of high quality. The series of higher order fractional quantum Hall states emanating from ν=1/2 is well developed and comparable in quality to that seen in

high mobility electron layers. Fractional states up to $\nu=6/13$ are visible. Given this initial data, one can expect that carbon-doped 2DHS will provide a novel platform for the study of correlations in two dimensions.

Finally, we discuss the transport properties of the lower density samples. As shown in Table 1, at $p=3.6\times10^{10}cm^{-2}$ and T=0.3K, no well width dependence of the mobility is observed. All three samples have a T=0.3K mobility of approximately 650,000cm$^2$/Vs. It is possible that the resulting perturbation to the valence band edge effective mass from non-parabolicity and quantum confinement is not as pronounced at $p=3.6\times10^{10}cm^{-2}$ as it is at $p=1\times10^{11}cm^{-2}$. At low density, the Fermi wave vector $k_F$ is small and the admixture of the light and heavy hole bands is not as significant close to the Brillouin zone center. The band edge mass would then remain small and relatively constant. Such an effect at low densities has been observed for 311A structures [15], and is consistent with the very high mobility (650,000cm$^2$/Vs at T=0.3K) observed for our low-density samples. We also note that the mobility of these low-density samples does not saturate at T=0.3K. When sample 10-24-03.2 was cooled to T=15mK, the mobility of the 2DHS increased to 800,000cm$^2$/Vs. It may be that all such low-density samples must be characterized at milliKelvin temperatures. This effort is underway. Figure 3 displays the longitudinal magnetoresistance of the 15nm quantum well at T=15mK. The quality of this structure is reflected in the strong quantization in the integer Hall regime and the strength of the fractional quantum Hall states at very small values of magnetic field. The $\nu=5/3$ state is well developed at B~0.9T and the $\nu=2/3$ state is fully quantized at B=2.25T. The quality of the magnetotransport at low 2D densities coupled with the lack of significant anisotropy indicates that (100) 2DHSs with density ~$10^{10}$cm$^{-2}$ will be useful for the exploration of 2D physics in the strongly interacting limit.

M. J. Manfra thanks Dan Tsui for useful discussions and for bringing the data of reference 15 to his attention.

**FIGURE CAPTIONS**:

**Figure 1**. Well width dependence of the T=0.3K mobility for several hole samples with density p=$10^{11}$cm$^{-2}$. All measurements are made in the Van der Pauw geometry, and the samples have been cooled to T=0.3K in the dark. The non-monotonic well width dependence of mobility is evident.

**Figure 2**. Magnetotransport data for sample 9-8-03.1 at T=20mK. Several of the principle fractional quantum Hall states are labeled.

**Figure 3**. Magnetotransport data for sample 12-24-03.2 at T=15mK. This sample is distinguished by strong quantization at very low magnetic fields ($\nu$=2/3 at B=2.25T).

**Table 1: Parameters of quantum well samples used in this study. All densities and mobilities refer to the values obtained at T=0.3K in the dark.**

| sample | Al mole fraction | width (nm) | p ($10^{11}$cm$^{-2}$) | μ ($10^5$cm$^2$/Vs) |
|---|---|---|---|---|
| 8-21-03.3 | 0.32 | single interface | 1.1 | 5.2 |
| 8-29-03.2 | 0.32 | 10 | 1.1 | 5.9 |
| 9-2-03.1 | 0.32 | 5 | 1.0 | 0.64 |
| 9-8-03.1 | 0.32 | 15 | 1.1 | 10 |
| 9-8-03.2 | 0.32 | 30 | 1.1 | 7.3 |
| 10-30-03.1 | 0.32 | 7 | 0.9 | 1.9 |
| 10-30-03.2 | 0.32 | 50 | 0.9 | 5.7 |
| 12-24-03.1 | 0.16 | 30 | 0.36 | 6.6 |
| 12-24-03.2 | 0.16 | 15 | 0.36 | 6.6 |
| 12-24-03.3 | 0.16 | 100 | 0.36 | 6.6 |

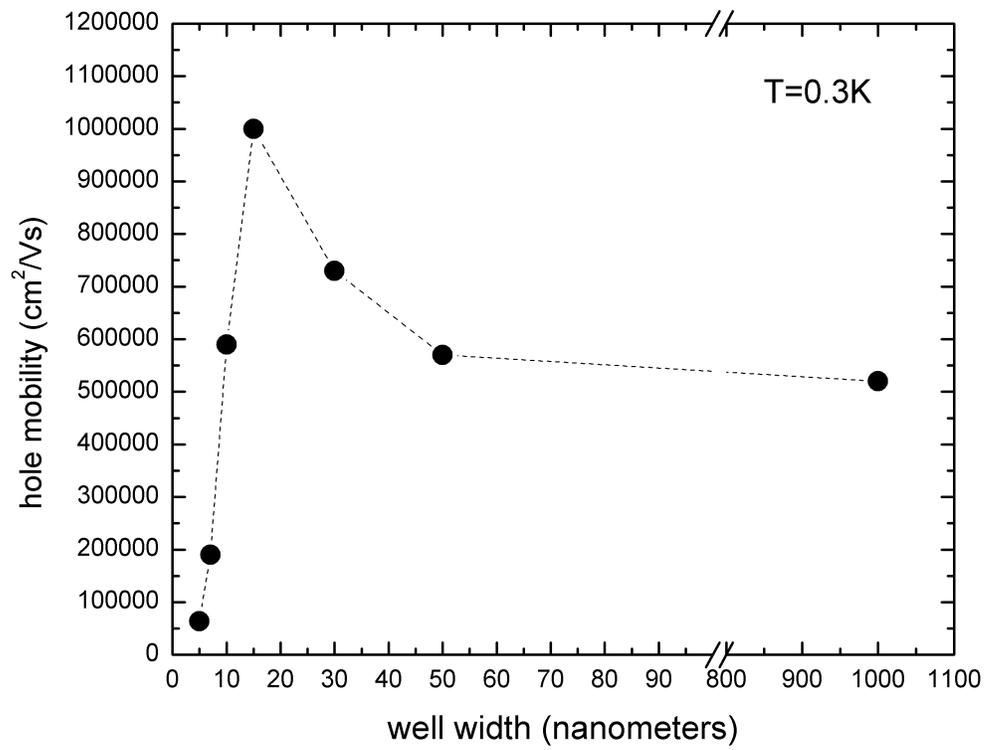

Figure 1. Manfra et al.

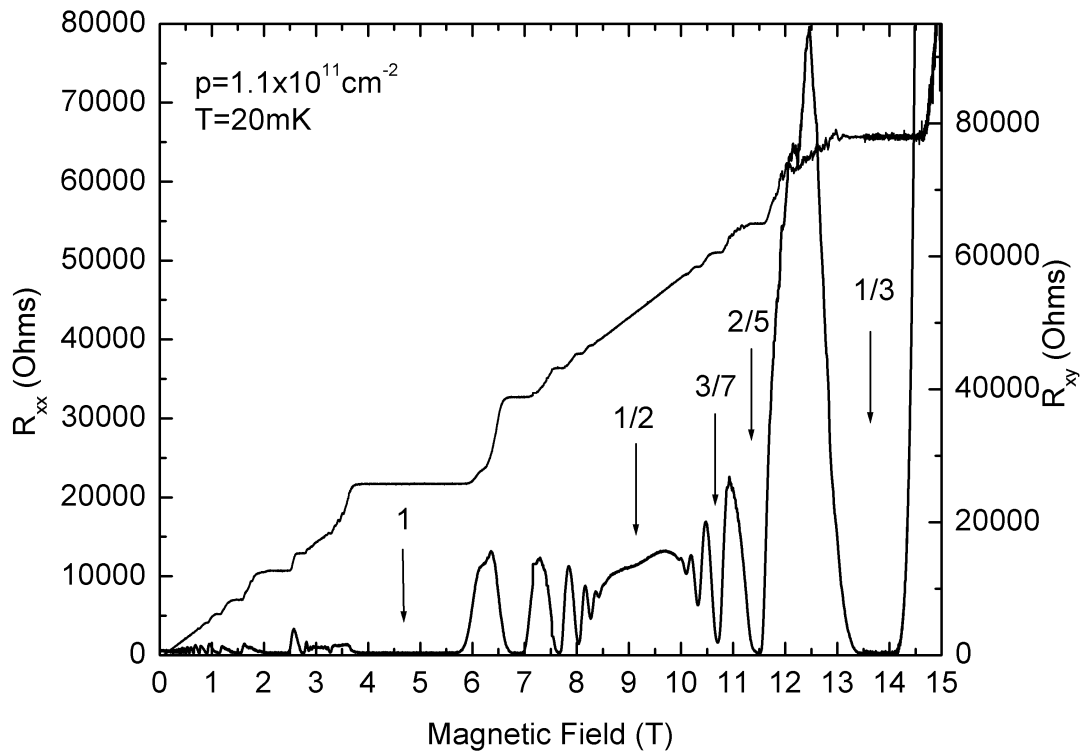

**Figure 2. Manfra et al.**

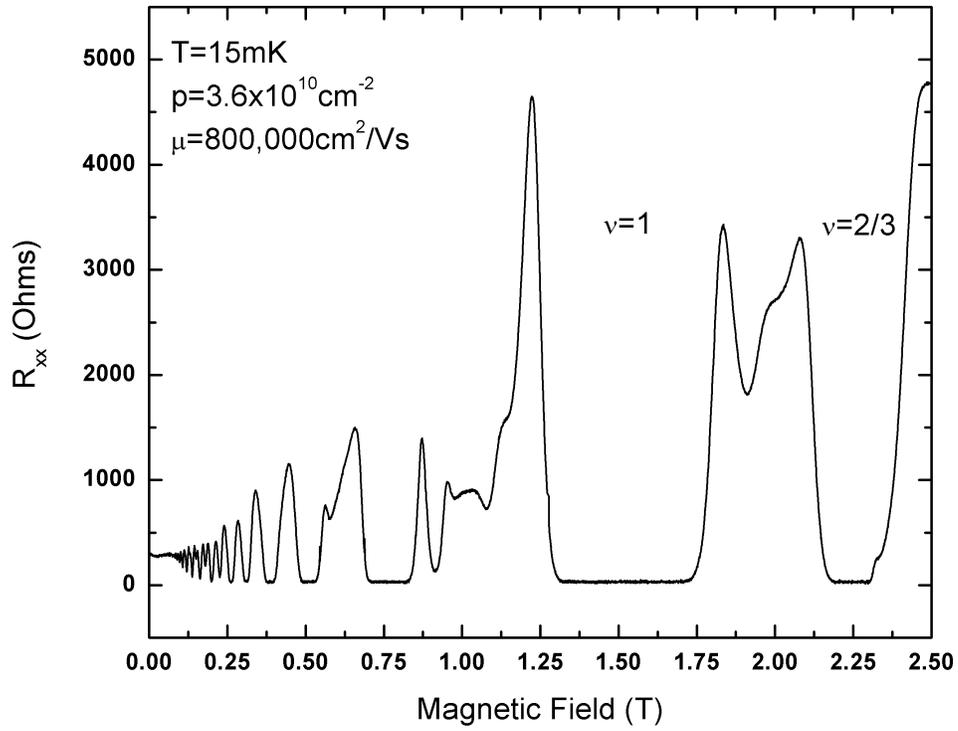

Figure 3. Manfra et al.